\documentclass{article}

\usepackage{PRIMEarxiv}

\usepackage[utf8]{inputenc} 
\usepackage{amsmath}
\usepackage{tabularx}
\usepackage[T1]{fontenc}    
\usepackage{hyperref}       
\usepackage{url}            
\usepackage{booktabs}       
\usepackage{amsfonts}       
\usepackage{nicefrac}       
\usepackage{microtype}      
\usepackage{lipsum}
\usepackage{fancyhdr}       
\usepackage{graphicx}       
\graphicspath{{media/}}     

\usepackage{multirow}
\usepackage{makecell}
\title{TaCNet: Temporal Audio Source Counting Network}

\author{
  Amirreza Ahmadnejad \\
Department of Electrical Engineering\\
	Sharif University of Technology\\
	Tehran, Iran, 11155-4365 \\
\texttt{amirreza.ahmadnejad@sharif.edu} \\
   \And
  Ahmad Mahmmodian Darviishani \\
  Department of Computer Science \\
	Institute for Advanced Studies in Basic Sciences\\
	Zanjan, Iran, 45137-66731 \\
  \texttt{AhmadMahmoodianDarvishani@gmail.com } \\
     \And
  Mohmmad Mehrdad Asadi \\
  Department of Electrical Engineering \\
	Sharif University of Technology\\
	Tehran, Iran, 11155-4365 \\
  \texttt{Mehrdadasadi3055@gmail.com } \\
       \And
  Sajjad Saffariyeh \\
  Department of Electrical Engineering \\
	Sharif University of Technology\\
	Tehran, Iran, 11155-4365 \\
  \texttt{sajjadsfryh@gmail.com} \\
         \And
  Pedram Yousefi \\
  Department of Electrical Engineering \\
	Sharif University of Technology\\
	Tehran, Iran, 11155-4365 \\
  \texttt{pedram1998yousefi@gmail.com } \\
           \And
  Emad Fatemizadeh \\
  Department of Electrical Engineering \\
	Sharif University of Technology\\
	Tehran, Iran, 11155-4365 \\
  \texttt{fatemizadeh@sharif.edu} \\
}

\begin{document}
\maketitle

\begin{abstract}
In this paper, we introduce the Temporal Audio Source Counting Network (TaCNet), an innovative architecture that addresses limitations in audio source counting tasks. TaCNet operates directly on raw audio inputs, eliminating complex preprocessing steps and simplifying the workflow. Notably, it excels in real-time speaker counting, even with truncated input windows.
Our extensive evaluation, conducted using the LibriCount dataset, underscores TaCNet’s exceptional performance, positioning it as a state-of-the-art solution for audio source counting tasks. With an average accuracy of 74.18 \% over 11 classes, TaCNet demonstrates its effectiveness across diverse scenarios, including applications involving Chinese and Persian languages. This cross-lingual adaptability highlights its versatility and potential impact.
\end{abstract}


\section{Introduction}
Inverse problems, characterized by the quest for latent 
causal variables from observable data \cite{cite1}, permeate diverse scientific disciplines. Concurrently, the domain of signal separation emerges as a foundational challenge within signal processing \cite{cite2}, accentuating its intricacy and ill-posed nature.
Signal separation, often considered a facet of inverse problems, encapsulates the pursuit of uncovering elemental constituents within complex amalgamated signals.

Prominently positioned within this discourse is the domain of Blind Signal Separation (BSS) \cite{cite2}, marked by dynamic progress and a profusion of algorithmic innovations. Nonethe-less, a defining limitation persiststhe predilection for a priori knowledge of source counts. Within the precincts of BSS, an array of methodologies thrives, excelling in rudimentary signal scenarios through precision models adept at source counting
\cite{cite5,cite6}. However, as the complexity escalates, as seen with intricate speech signals, conventional methodologies falter.
It is pertinent to underscore that while some audio source separation models present integrated solutions that holistically address both separation and counting for audio signals \cite{cite7,cite8,cite9,cite10,cite11}, the incorporation of a dedicated speaker counting module remains an avenue ripe for exploration.

Transcending the realm of audio source separation, the role of speaker enumeration finds broader resonance. Applications including localization \cite{cite12}, diarization \cite{cite13}, and identification
\cite{cite14} underscore the potency of speaker counting to streamline computations and bolster model versatility. Clearly, foundational to sound processing endeavors is the cognizance of speaker counts.

In the contemporary landscape,
deep
neural
networks(DNNs) have ascended to the forefront, offering
formidable prowess in resolving multifaceted audio processing
challenges \cite{cite15,cite16,cite17,cite18,cite19,cite20,cite21}. This trajectory extends to the realm of
audio source counting, evident through scholarly discourse exploring neural network-driven solutions, as comprehensively
expounded in Section 2.

Yet, the challenge of feature extraction remains an enduring
it dilemma within learning paradigms. Approaches span the spectrum from harnessing raw audio inputs \cite{cite16} to effectuating
transitions into the frequency domain \cite{cite15}. Historical reliance on meticulously crafted features, typified by the Mel-filterbank
\cite{c22} for counting tasks, has demonstrated utility. However, the efficacy wanes with escalating source counts.

The crux of innovation manifests in our model's inception,where inspiration is gleaned from the Mel-filterbank paradigm.
Synonymous with its essence, our architectural design features filters, down-sampling, and sequential comparison. In particular, filters serve to emphasize salient spectral components while down-sampling systematically reduces temporal resolution. Subsequent comparison engenders discernment of distinctive features, orchestrating a transformative process akin to the transformative attributes of Mel filterbanks. A distinctive hallmark of our model lies in the endowment of learnable attributes to each parameter, endowing it with adaptability and refinement potential. This holistic fusion of elements nurtures a model poised to evolve and adapt to the complexities inherent in speaker counting.

We conceptualize audio speaker counting as a classification task, where our model is formulated to operate on single-channel audio sources. One notable attribute of our model is its capacity for efficient real-time counting of audio sources, facilitated by its input frame size and consequent low latency processing.

Our empirical investigations attest to the efficacy of leveraging features extracted from raw audio, eclipsing the conventional reliance on handcrafted counterparts. These advancements culminate in a state-of-the-art model demonstrated on the LibriCount dataset, encompassing scenarios with speaker counts spanning from 0 to 10. Furthermore, our model's robustness is underscored through transfer learning experiments on Chinese and Persian datasets, attaining commendable outcomes sans any compromise in accuracy.
Thus, this paper unfolds in a structured manner as follows:
Section II: Explores the landscape of related works and delineates diverse approaches undertaken in audio speaker estimation. Section III: Unveils the architecture of our model, elucidating its intricate expression and design. Section IV:
Engages in a comprehensive discussion of varied deep neural network architectures deployed across distinct components of our model.Section V: Presents a holistic exposition of experiments conducted, encompassing diverse features and classifiers. This section also entails a comparative analysis between our model and alternative counting paradigms, accompanied by a presentation of other implementations. Section VI: Concludes the paper, encapsulating the essence of our findings and illuminating avenues for future exploration.

\section{Related Works}
This section delves into prior endeavors concerning single-channel audio source counting. Each instance presents distinct merits and limitations, with a common emphasis on the choice of features and input modalities for the counting model.
Pioneering work by \cite{c25} explored the nexus between modulation index and the total speaker count in audio. The modulation index function was leveraged to estimate the number of speakers from modulated input signals, proficiently discerning up to 8 simultaneous speakers in TIMIT dataset samples \cite{cite26}.

Subsequent forays, such as \cite{c27}, established a correlation between the number of speakers in a single-channel mix and a specific Mel-filterbanks coefficient. Employing regression methodology, a polynomial function emerged as the designated counting model.

In \cite{c28}, a clustering algorithm found application in accumulating MFCCs from an equitably distributed speaker pool.
The cosine similarity formed the bedrock of the objective function, fostering a clustering algorithm that juxtaposed MCC features for optimal grouping.

In lieu of Mel-filterbanks, \cite{c29} and \cite{c30} ventured into the Dynamic Time Warping (DTW) domain, employing it to measure audio similarity. This approach identifies audio samples with matching speaker counts, necessitating a reference audio for comparison. In \cite{c29} specifically gauged human-versus-machine speaker counting, revealing machines' superior per-formance, especially for short audio durations.

The advent of deep learning surfaced with \cite{c31}, juxtaposing regression and classification for speaker counting problems.
Both pathways featured DNNs armed with bidirectional Long short-term memory (LSTM) layers. A suite of handcrafted features, including Mel-filterbanks, MFCC, STFT, and LOG-STFT, were evaluated. STFT-based features emerged as optimal for classification, culminating in an advanced record achieved via diverse DNN architectures in \cite{c32}, where Con where Convolutional Recurrent Neural Networks (CRNN) emerged as a potent classifier.

Following these milestones, subsequent efforts turned toward diverse DNN models, yet the central input feature often remained confined to Mel-filterbanks or STFT coefficients.
The influx of CNNs as feature extractors finds resonance in the transition from image classification challenges \cite{c33} to audio contexts. CNNs have notably excelled in extracting robust features from raw audio, as demonstrated in speech separation endeavors \cite{cite16}.

Despite these advancements, the pursuit of an adaptable feature universally applicable to varied audio processing tasks persists. Recent years have witnessed several innovative propositions to surmount this challenge, illuminating novel pathways for speaker counting feature extraction. The first stems from \cite{c34}, which advocates Transfer Learning of the SincNet Model's bottleneck layer \cite{c35} as an extractor of features.
The second, detailed in \cite{c36}, integrates an attention mechanism \cite{c37} to discern optimal segments within the Log Mel-filterBank for feature extraction. The final, and most closely aligned with our work, is discussed in \cite{c38}, which undertakes speech detection and counting from raw audio, exemplifying a pioneering endeavor to derive novel features instead of relying on handcrafted ones.

It's noteworthy that \cite{c38}, while innovative, presents certain limitations, which our work aims to address:
Firstly, it solely evaluates audio scenarios featuring up to 4 speakers. This highlights a critical aspect - the necessity of new features that perform robustly as the speaker count escalates.
Secondly, as detailed in the subsequent section, distinct alternatives to utilizing pristine convolution layers as primary feature extractors have been proposed. For instance, SinNet \cite{c35} introduces filters with distinctive sink shapes in early layers. Importantly, \cite{c38} predominantly accounts for audio samples with up to 4 speakers.

\section{TaCNet Model}
Let $x[n] \in \mathbb{R}^T$  denote a digitized audio signal within the time domain, sampled at a rate of F. The individual audio sources are represented as sin, where the count of sources can range from 1 to N. Each $s_i [n]$ signal exists in the time domain, sampled at the rate $F_s$. Operating on the principle of summation, it is postulated that the individual source signals must possess an equal length for coherent summation. To adhere to this, the length of each speaker signal is set at $T$.
Consequently, the aggregated audio signal, denoted as $x[n]$, can be expressed as:

\begin{equation}
    x[n] = \sum_{i=1}^{N} s_i [n]
\end{equation}

However, it's important to consider that some speakers may remain silent for certain periods within the audio signal, represented as 'inactive segments' within the waveform $x[n]$. This occurs when there are variations in speaker activity over time. For example, during one portion of the audio, there might be four active speakers, while in another part, there may be only three, and subsequently, two active speakers. This variability in the number of active speakers necessitates dividing the audio signal $x[n]$ into discrete chunks, each containing the exclusive audio contribution of a single active source.

The function responsible for windowing the audio signal $x[n]$ can be expressed as:
\begin{equation}
    w_i [n] = u[n-a_i]-u[n-b_i]
\end{equation}

In the above equation, $u[\cdot]$ represents a step function, and it is implicitly assumed that $b > a$. The subscripts $i$ for $a$ and $b$ indicate the start and end points of the window, respectively.
For instance, the first window is defined as $w_1 [n] = u[n]- u[n-b_1]$, and the second window is $w_2 [n] = u[n-a_1]- u[n-b_2]$.

The length of this window is held constant in this study, although it could be made dynamic based on the input signal.
Exploration of this adaptive approach is left for future research. For the purposes of this work, the window length is set at 10 ms. This choice not only supports real-time processing but also considers that excessively shorter window lengths are unlikely to provide advantages, as such rapid speech rates are typically limited to artificial speakers. Nevertheless, an investigation into the impact of this fixed window size on output responses will be conducted.

Subsequent to the application of the window function $w[n]$ to each segment of the input signal, a distinct label is assigned to each chunk, denoted as $y$. Hence, the pair
$(x_i = w_i [n] \times x[n], y_i)$ constitutes a sample within the training
set. Consequently, the problem of audio source counting is formulated as follows:

\begin{equation}
    y_i = f_{\theta}(x_i)
\end{equation}

\begin{figure*}[ht!]
\begin{centering}
\includegraphics[width=\linewidth]{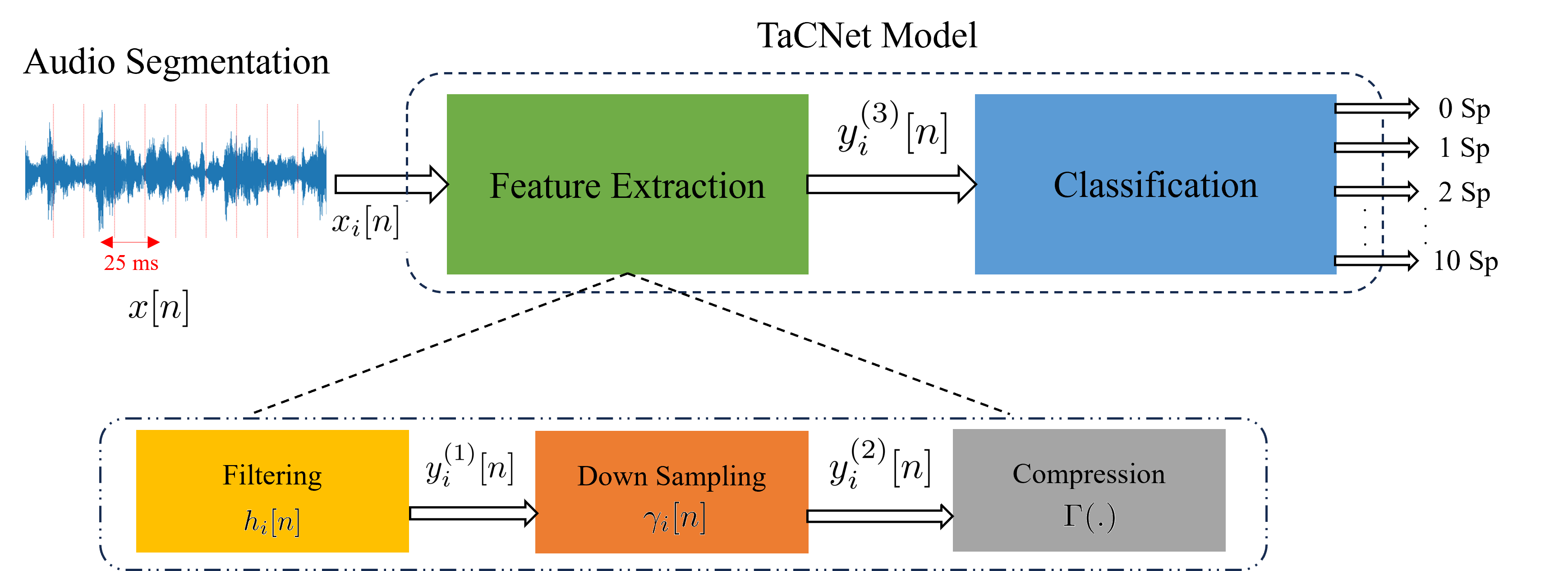}
\par\end{centering}
\caption{Overview of the TaCNet Model:
The TaCNet model operates through a structured sequence of operations. Initially, the audio file is segmented into a defined number of partitions. Subsequently, these partitions undergo feature extraction through a dedicated block comprising three stages: filtering, downsampling, and compression. The extracted features are then directed to the classification block, which ultimately serves to ascertain the number of speakers present in the input audio.}
\end{figure*}

The function $f_{\theta} (.)$ signifies the objective function that establishes the connection between each chunk and its corresponding label. This transformation is realized through a DNN, denoted as $f_{\theta}$ with $\theta$ representing the model parameters.

Reference \cite{c31} empirically determined that the optimal solution for Equation 3 involves classification. Accordingly, we structure this problem using two distinct modules:
 - The first module aims to learn features from the audio data.
Generally, this module can be expressed as $g_{\theta} : \mathbb{R}^{T} \rightarrow \mathbb{R}^{M\times N}$, where it maps the 1D time-domain raw audio to a 2D feature space. Within this mapping, $M$ signifies the temporal frames, $N$ represents the number of feature channels, and $\theta$ encompasses the parameters of the entire feature extractor.

Subsequently, these 2D feature vectors are input to the classification module, which can be denoted as $h_{\phi}$. Ultimately, both of these modules are trained in an end-to-end manner by minimizing the following cost function:

\begin{equation}
    C= argmin_{\theta,\phi} \sum_{j=1}^{D} \sum_{i=1}^{K} L(h_{\phi}(g_\theta(x_{i,j},y_{ij})))
\end{equation}

In the above equation, the subscript $j$ pertains to dataset instances, $D$ signifies the number of data points, the subscript $i$ denotes individual chunks, and due to the fixed window size, each audio source is divided into $K$ parts.

Regarding the architectural design of the first module, references \cite{c22} and \cite{c39} have explored optimal feature extraction from raw audio for audio classification. Their focus centers on the Mel-filterbank features, which encompass three key components: Filtering, Downsampling, and Compression.

In the Mel representation, the signal $x(t)$ is initially subjected to band-pass filtering followed by a non-linear operation. This function is performed at the same sampling rate as the input signal. Subsequently, pooling is applied to the signal to accommodate its reduced resolution, concluding with the application of a compression unit to reduce the dynamic range.
An important principle presented by \cite{c23} is the separation of each level within the Mel representation, with learnable hyperparameters linked to the input signals. This permits the creation of a fully learnable front-end module controlled by a relatively small number of parameters. The entire process is illustrated in Figure 1.

The subsequent sections will elaborate on each constituent part of the model, encompassing both the feature extraction and classifier modules.

\subsection{Filtering}

The initial stage involves convolving the audio signal $xIn]$ with a bank of complex-valued filters $h_i [n]$ where $i=1, 2, .., N$. Subsequently, a absolute squared operator is applied to yield real-valued outputs. The convolution stride is set to one, maintaining the size consistency between input and output.
The operation can be represented as:

\begin{equation}
   y_i ^{(1)} [n] = |x[n]*h_i [n]|^2 \in \mathbb{R}^T , i=1,2, .., N
\end{equation}

The superscript $(1)$ denotes the output of the first module. An alternative method for computing Equation 5 is proposed by
\cite{c23}, which is left for readers to explore. For selecting $h_i[n]$, various transformation functions can be employed. Notably, the innovation by \cite{c23} involves using 1-D Gabor filters instead of normalized 1D-convolution \cite{c39} or the Sinc function \cite{c35}.
Gabor filters offer the advantage of possessing the same representation in both time and frequency domains, enhancing model interpretability. Gabor filters are determined by two key factors: center frequencies $\mu_i$ and bandwidths $\sigma_i$
where $i = 1, ...,N$.
The transformation function for the first component is expressed as:
\begin{equation}
   h_i [n] = e^{j2\pi \mu_i n} \frac{1}{\sqrt{2\pi} \sigma_i} e^{-\frac{n^2}{2\sigma_i ^2}}, i=1, 2, .., N
\end{equation}

In this equation, $j=\sqrt{-1}$, $n$ ranges from $n = -\frac{W}{2}$ to $n=\frac{W}{2}$
encompassing all filters defined within this interval.

\subsection{Down Sampling}
Following Filtering, the output matches the size of the input audio signal. Downsampling aims to reduce signal resolution, akin to extracting Mel-filterbank features through Short-Time Fourier Transform (STFT).

Previous works in speech recognition applied various methods for downsampling, such as max-pooling \cite{c35}, low-pass filtering \cite{c40}, or average-pooling \cite{c41}. In \cite{c23}, Gaussian low-pass filtering is employed on each output channel post the filtering module. This approach is efficient as different bandwidths can characterize each channel within the learnable model. Moreover, the Gaussian filter is a specific case of Gabor filters with a center frequency of 0 and a learnable bandwidth.
This introduces only a few learnable parameters to the overall model.
Considering that N filters were applied in the previous step, the low-pass filters in this module can be expressed as:
\begin{equation}
   y_i ^{(2)}= y_i ^{(1)}*\gamma_i [n] = y_i ^{(1)} * (\frac{1}{\sqrt{2\pi} \sigma_i} e^{-\frac{n^2}{2\sigma_i ^2}})
\end{equation}
Here again, $i=1, 2, .., N$ and $n$ ranges from $n = -\frac{W}{2}$ to $n=\frac{W}{2}$.

\subsection{Compression}
For hand-crafted features like Mel-filterbank, the output time-frequency features typically undergo a nonlinear operation (logarithm) to simulate human perception of volume.
However, this approach compresses all frequency bins uniformly. An alternative to the logarithm is Per-Channel Energy Normalization(PCEN) \cite{c42}, which combines logarithmic and mean-variance normalization as follows:
\begin{equation}
   y_i ^{(3)}= \Gamma (y_i ^{(2)}) = (\frac{y_i ^{(2)}[n]}{(\epsilon+y_i ^{*})^{\alpha_i}}+\delta_i)^{r_i} - \delta_i^{r_i}
\end{equation}
Here, $\Gamma(.)$ is the nonlinear learable function (PCEN), $n = 1,..., M$
represents the time step, and $i = 1,..., N$, corresponds to the
channel number. $y_i ^{(2)}[n]$ is normalized with respect to past values  $y_i ^{*}[n] = (1-s) y_i ^{*}[n-1] + sy_i ^{(2)}[n]$, controlled by
coefficients $s$ and $\alpha_i$. Additionally, $\epsilon$ is a constant to prevent division by zero, and $\delta_i$; is an offset. The comparison process involves the exponent $r_i$, typically within the range $r_i \in [0,1]$.

In summary, the feature extraction module incorporates a sequence of 1D-convolution with Gabor kernels, Gaussian low-pass pooling, and the nonlinear function.

\subsection{Classifier}
Upon completion of the feature extraction module, the original input $x[n]$ in the time domain with length $T$ transforms into an $N \times M$ feature matrix in the time-frequency domain.
As with previous stages, the resulting output is denoted as $y_i ^{(3)}[n]$. This feature matrix is then fed into the classifier module and trained using supervised multi-label classification.
As mentioned earlier, if we denote the entire feature extraction module as $g_{\theta}(.)$ and the classifier as $h_{\phi}(.)$, the audio source counting process can be expressed as:
\begin{equation}
   \hat{y} = h_{\phi} (g_{\theta}(x[n]))
\end{equation}
When $x[n]$ is input into the classifier, the network generates posterior output probabilities for $\zeta + 1$ classes where $\zeta$ is the maximum number of classes, where $\hat{y}$ is a vector of length $\zeta + 1$. The inclusion of the additional class accounts for the detection of no speaker or noise. As mentioned in \cite{c31}, while classification for audio counting yields superior results compared to regression, it presents two limitations:

Firstly, there exists no intrinsic meaningful relationship between distinct classes. Consider an audio file with three speakers. Counting the number of sources can be approached as either the sum of three single-speaker sounds or the sum of a two-speaker sound and a single-speaker sound. In the classification model for source counting, this relationship is disregarded, and its exploration is reserved for future studies.

Secondly, the potential count $\zeta$ is predetermined prior to model training and testing during classification. This implies that even before the model is trained or tested, one can predict that counting accuracy decreases as the number of classes increases. However, a sudden improvement is observed for the last class. This phenomenon underscores the classifier's awareness of $\zeta$.

\section{TaCNet Architecture}
The architecture utilized for the feature extractor involves a standard convolutional layer, followed by a customized pooling layer, and culminating with a specific activation function. In the initial convolutional layer, N = 40 filters are
employed, each comprising W = 401 coefficients. Figure 2 provides a concise overview of the entire architecture.

\begin{figure*}[ht!]
\begin{centering}
\includegraphics[width=\linewidth]{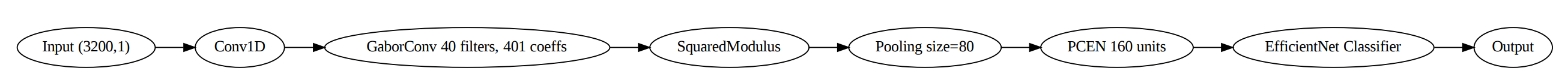}
\par\end{centering}
\caption{The general outline of the model architecture. Initially, the audio signal, which is sampled over a 25-millisecond window and has a length of 1*3200, is inputted into the 1D convolution layer. Subsequently, Gabor filters are applied to the signal, and it is then fed into the classifier, as explained in the architecture.}
\end{figure*}

For the classifier module, we employed three distinct archi-tectures: Efficient-Net \cite{c42}, Pretrained Audio Neural Networks for Audio Pattern Recognition (PANN) \cite{cite20}, and CNN-14.
Notably, \cite{c32} conducted an assessment of diverse classifier architectures and identified CRNN as the most effective. Thus, we examine these four distinct architectures individually and present the outcomes.
Among the considered classifiers, it is worth noting that the most promising performance is observed with the Efficient-Net architecture. This architecture delivers the most favorable results in terms of the audio source counting task.

\section{Experiments}
\subsection{General Information}
We employed the LibriCount dataset introduced in \cite{c22}.
This dataset is derived from Librispeech and comprises approximately 8 hours of audio recordings. LibriCount spans the range from zero speakers to ten speakers, with each label having an equal representation of 572 data instances. The initial step involves segmenting each audio file into 5-second intervals, utilizing a sample rate of 16000. For preprocessing, we explored window sizes ranging from 10 ms to 40 ms, with a step of 5 ms adopted in separate approaches. The labels assigned are generated based on speaker activity. Overall, this process yields approximately 1,144,000 data instances for the case of a 25 ms window. These instances are then partitioned into training, validation, and testing sets.

An essential aspect to highlight is the application of the Mode function post-segmentation. Figure 3 illustrates a sample from the dataset labeled as a 4-speaker source. It might prompt skepticism if a chunk of 500 samples is chosen, where half of the audio corresponds to label A and the other half to label B. 

\begin{figure}[h]
    \centering
    \includegraphics[scale=0.3]{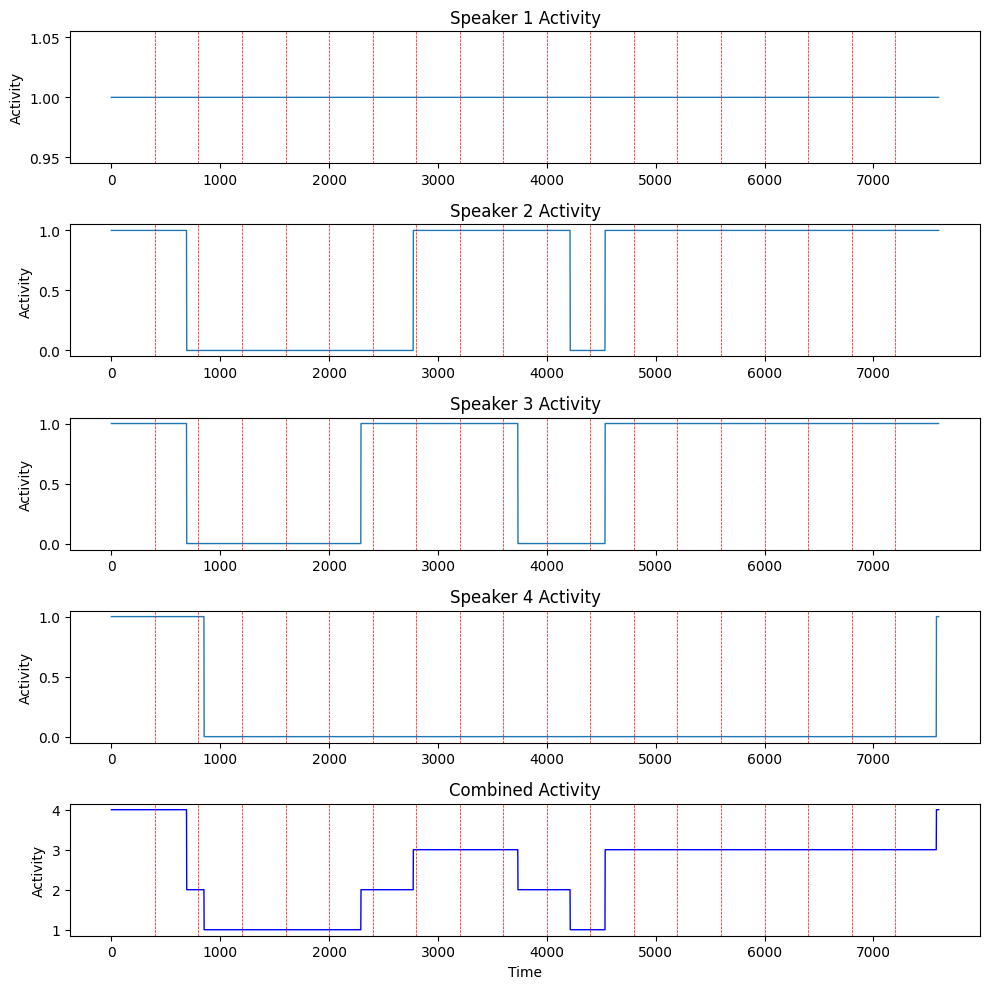}
    \caption{Within the LibriCount dataset, we present an illustrative instance featuring four distinct speakers engaged in discourse. The initial segment focuses on the uninterrupted vocal presence of the first speaker throughout the entire duration of the audio recording. The subsequent segments are dedicated to the second, third, and fourth speakers, respectively, each exhibiting periods of vocal inactivity. The final segment portrays the composite (overlapping) representation of these speaker activities, encapsulating a heterogeneous ensemble of speakers distributed across various temporal segments within the audio file.}
    \label{fig:image-label}
\end{figure}

To alleviate this concern, it's important to note that the window size is sufficiently small, rendering such a situation highly unlikely. The co-occurrence of two distinct labels with such disparities within this limited window size is unusual.
Furthermore, the likelihood of encountering this issue scales when considering the entirety of the LibriCount dataset.

Following a comprehensive evaluation of various window sizes, a window size of 25 ms emerged as optimal for preprocessing, yielding lower Mean Absolute Error (MAE), as demonstrated in Figure 4. Consequently, the model can effectively operate in an online manner. It's worth highlighting that dynamic window sizing, tailored to the characteristics of each audio input, could serve as another trainable parameter, an avenue reserved for future research.

\begin{figure}[h]
    \centering
    \includegraphics[scale=0.6]{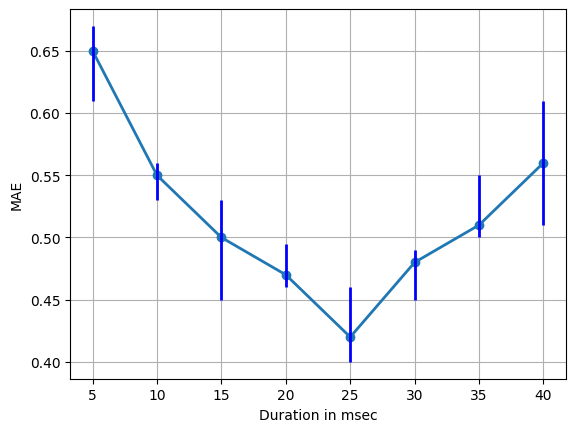}
    \caption{Mean Absolute Error (MAE) observed across various window sizes exhibits a notable pattern. Notably, there is an initial decrease in the MAE, followed by a subsequent increase. It is worth highlighting that the minimum error is associated with a window size of 25 milliseconds.}
    \label{fig:image-label}
\end{figure}

The training procedure was conducted using Google Colab, with a duration of approximately 6 hours. The Graphics Processing Unit (GPU) employed was the NVIDIA Tesla K80. Given the multilabel classification nature of the audio source counting task, we adopted accuracy as the primary evaluation metric. Additionally, the confusion matrix was employed to visualize the classifier's output for each class, which is depicted in Figure 5.

\begin{figure}[h]
    \centering
    \includegraphics[scale=0.4]{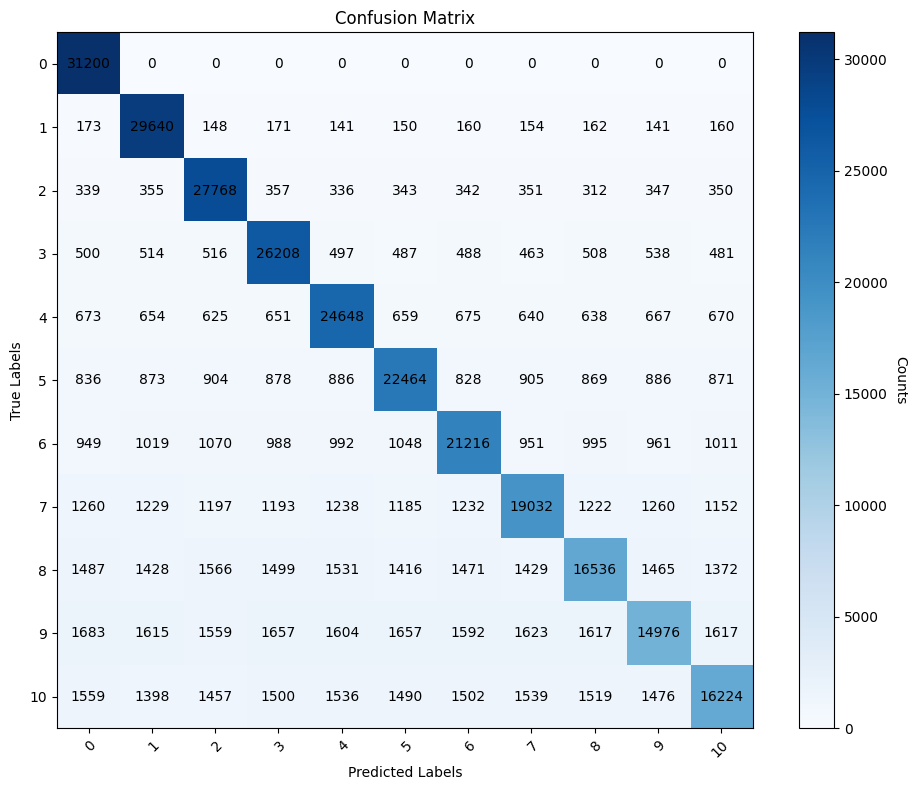}
    \caption{The confusion matrix illustrates the performance of the speaker counting system on the test set. The diagonal elements represent the frequency of correct counts, while the off-diagonal elements indicate erroneous counts. Specifically, the main diagonal contains the number of test segments for which the estimated speaker count matched the true number of speakers. The off-diagonal elements show the number of instances where the system incorrectly estimated the speaker count.}
    \label{fig:image-label}
\end{figure}

An observed trend indicates that as the number of speakers increases, the classifier's performance diminishes. However, if the primary goal involves counting speakers up to four, it may not be justifiable to develop a new model, as the existing learned counting model demonstrates robust performance.

\begin{table*}[t] 
  \caption{a comparison of speaker counting accuracy across different models}
  \centering
  \begin{tabular}{|c|c|c|c|c|c|c|c|c|c|c|c|}
    \hline
    Model & 0 Sp & 1 Sp & 2 Sp& 3 Sp& 4 Sp& 5 Sp& 6 Sp& 7 Sp& 8 Sp& 9 Sp& 10 Sp\\
    \hline
    Stoter et al.\cite{c31} & 100 & 92 & 86& 74& 67& 41& 37& 31& 45& 55& 49\\
    \hline
        Wang et al.\cite{c34} & - & 99 & 85& 81& 56& 68& 40& 41& 25& 29& 68\\
            \hline
        Stoter et al.\cite{c32} & 98 & 99 & 90& 81& 69& 59& 55& 39& 35& 38& 68\\
                    \hline
        TaCNet & 100 & 95 & 89& 84& 79& 72& 68& 61& 53& 48& 71\\
    \Xhline{2pt}
        Yousefi et al.\cite{c36} & - & 100 & 91& 75& 82& -& -& -& -& -& -\\
    \hline
        Zhang et al.\cite{c38} & - & 94 & 52& 36& 83& -& -& -& -& -& -\\
    \hline
        Andrei et al.\cite{c30} & - & 88 & 80& 74& -& -& -& -& -& -& -\\
    \hline
  \end{tabular}
\end{table*}

\subsection{Comparison between Counting models}
For comparative analysis of different models, we contrast our proposed TaCNET model against other notable approaches tested on the LibriCount dataset, including \cite{c32,c34,c36,c38}. Additionally, as illustrated in Table 1, alternative approaches highlighted in the previous works section are listed. The models above the bold line were trained and evaluated on the full LibriCount . The models below the line were trained on separate datasets and tested on a limited range of speakers. 
As demonstrated in Table 1, our proposed model exhibits superior performance, outperforming the best results attained by other models. The effectiveness of handcrafted features diminishes as the number of speakers increases and their patterns become more intricate. However, our feature extractor excels in learning superior features from raw audio inputs through the utilization of learnable features and small window sizes. This capability enables our model to discern complex patterns and achieve enhanced performance.

\subsection{Transfer Learning on various Languages}
A significant challenge within the realm of resource separation and counting pertains to the impact of language variation during model training. However, a model designed with broad generalizability, trained regardless of the language characteristics of input data, should ideally exhibit robust performance.
In an effort to enhance the inclusiveness of the TaCNet model, we conducted tests involving Chinese and Farsi languages during the testing phase. These tests were conducted using the same window size of 25 milliseconds, and the model's training weights remained unchanged.

The output results of these language-specific tests are presented in Figure 6. As depicted in the figure, the accuracy output results across three distinct datasets, each in different languages, exhibit a consistent pattern with minimal deviation.
This observation supports the assertion that TaCNet is capable of performing effectively across diverse languages.

\begin{figure}[h]
    \centering
    \includegraphics[scale=0.6]{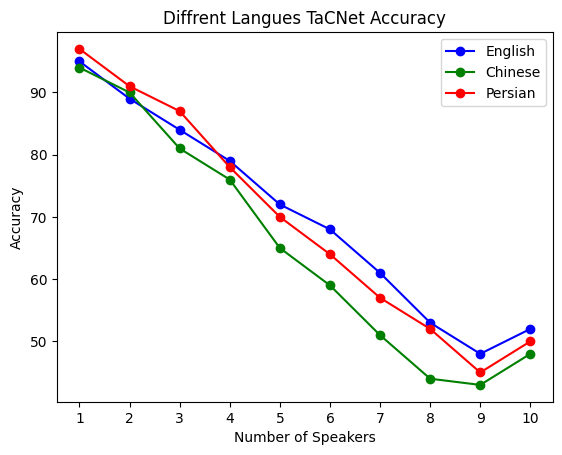}
    \caption{Transfer learning was conducted on the TaCNet model across multiple languages during the testing phase. The results reveal minimal disparity in output percentages among the models. Much like the original model, all of the models exhibited a decline in classification accuracy until the final class, with an eventual increase observed in the last class. This behavior can be attributed to the inherent characteristics of the classification task.}
    \label{fig:image-label}
\end{figure}

\section{Conclusion}
In this work, we endeavor to find better feature than handcrafted ones for audio source counting. Counting sources can  be helpful in different audio processing problems, especially audio separation. Our model learns suitable features and classifies them as a counting model. Due to the small window size at preprocessing level, our model can perform online processing but find an appropriate dynamic window size postponed for further work. Ultimately, we check the model efficiency by testing it on the Libricount dataset and achieving the best audio source counting in return for different models. As a perspective for further work, this model can be used as an isolated module before separation models, and the counting output can help the separation be done better.

\section{Acknowledgments}

I would like to express my gratitude to OpenAI for the invaluable assistance provided by ChatGPT \cite{c45} during the revision and refinement of this paper. ChatGPT played a significant role in helping to improve the clarity and coherence of the text. Its capabilities in natural language understanding and generation were instrumental in generating suggestions and alternative phrasings that greatly contributed to the overall quality of this work.

\bibliographystyle{unsrt}
\bibliography{sample}

\end{document}